# Concrete Attribute-Based Encryption Scheme with Verifiable Outsourced Decryption

Charan[1], K Dinesh Kumar[2] , D Arun Kumar Reddy[3]

[1]P.G Scholar, [2]Assistant Professor, [3]Associate Professor

[1,2,3]CSE, Sir Vishveshwariah institute of science and technology, Madanapalle, India

**Abstract:**

As more sensitive data is shared and stored by third-party sites on the internet, there will be a need to encrypt data stored at these sites. One drawback of encrypting data is that it can be selectively shared only at a coarse-grained level. Attribute based encryption is a public-key-based one-to-many encryption that allows users to encrypt and decrypt data based on user attributes. A promising application of ABE is flexible access control of encrypted data stored in the cloud using  access policies and ascribed attributes associated with private keys and ciphertexts. This functionality comes at a cost. In typical implementation, the size of the ciphertext is proportional to the number of attributes associated with it and the decryption time is proportional to the number of attributes used during decryption. Specially, many practical ABE implementations require one pairing operation per attribute used during decryption. One of the main efficiency drawbacks of the existing ABE schemes is that decryption involves expensive pairing operations and the number of such operations grows with the complexity of the access policy. Recently green *et al.* proposed an ABE system with outsourced decryption that largely eliminates the decryption overhead for users. tn such a system a user provides an untrusted server, say a cloud to translate any ABE ciphertext satisfied by that user's attributes or access policy into a simple ciphertext and it only incurs a small computational overhead for the users to recover the plaintext from the transformed ciphertext. Security of an ABE system with outsourced decryption ensures that an adversary will not be able to learn anything about the encrypted message; however it does not guarantee the correctness of the transformation done by the cloud. In this paper we consider a new requirement of ABE with outsourced decryption: verifiability. Informally, verifiability guarantees that a user can effectively check if the transformation is done correctly. We prove that our new scheme is both secure and verifiable without relying on random oracles. Finally, we show an implementation scheme and result of performance measurements, which indicates a significant reduction on computing resources imposed on users.

I. Introduction

There is a trend for sensitive user data to be stored by third parties on the internet. For example personal email, data and personal preferences are stored on web portal sites such as google and yahoo. The attack correlation center, dshield.org, presents aggregated views of attacks on the internet, but stores intrusion reports individually submitted by users. Given the variety, amount and the importance of information stored at these sites, there is cause for concern that personal data will be compromised. In distributed settings with untrusted servers, such as the cloud many applications need mechanisms for complex access control over encrypted data, sahai and waters[1] addressed this issue by introducing the notion of attribute based encryption. ABE is a new public key based one-to-many encryption that enables access control over encrypted data using access policies and ascribed attributes associated with private keys and cipher texts the cryptosystem of sahai and waters allowed for decryption when at least k attributes overlapped between a ciphertext and a private key. While this primitive was shown to be useful for error-tolerant encryption with biometrics the lack of expressebility seems to limit its applicability to larger systems. There are two kinds of ABE schemes: key-policy ABE(KP-ABE) [2]-[7] and ciphertext-policy ABE (CP-ABE) [8], [9], [5], [6]. In a CP-ABE scheme, every ciphertext is associated with an access policy on attributes and every user's private key is associated with a set of attributes. A user is able to decrypt a ciphertext only if the set of attributes associated with the user's private key satisfies the access policy associated with the ciphertext. In a KP-ABE scheme, the roles of an attribute set and an access policy are swapped from what we described for CP-ABE: attributes sets are used to annotate the ciphertexts and access policies over these attributes are associated with user's private keys. One of the main efficiency drawbacks of the most existing ABE schemes is that decryption is expensive for resource-limited devices due to pairing operations and the number of pairing operations required to decrypt a ciphertext grows with the complexity of the access policy. At the cost of security only proven in a weak model there exist several expressive ABE schemes[10], [11] where the decryption algorithm only requires a constant number of pairing computations. Green *et al[12]* proposed a remedy to this problem by introducing the notion of ABE with outsourced decryption, which largely eliminates the decryption overhead for users. Based on existing ABE schemes Green *et al* [12] also presented concrete ABE schemes with outsourced decryption. tn these schemes a user provides an untrusted server, say a proxy operated by a cloud service provider, with a transformation key TK that allows the latter to translate any ABE ciphertext CT satisfied by that user's attributes or access policy into a simple ciphertext CT and it only incurs a small overhead for the user to recover the plaintext form the transformed ciphertext CT. the security property of the ABE scheme with outsourced decryption guarantees that an adversary be not able to learn anything about the encrypted message; however the scheme provides no guarantee on the correctness of the transformation done by the cloud server.

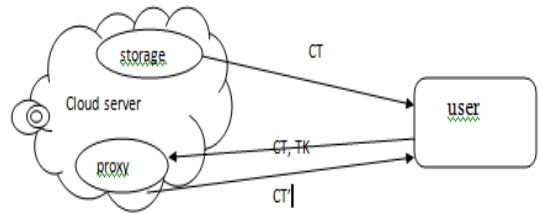

*Fig: ABE system with outsourced decryption.*

Consider a cloud based electronic medical record system in which patients medical records are protected using ABE schemes with outsourced decryption and are stored in the cloud. Inorder to efficiently access patients medical records on her mobile phone a doctor generates and delegates a transformation key to a proxy in the cloud for outsourced decryption.

Given a transformed ciphertext from the proxy, the doctor can read a patient's medical record by just performing a simple step of computation. If no verification of the correctness of the transformation is guaranteed, however the system might run into the following two problems: 1) for the purpose of saving computing cost, the proxy could return a medical record transformed previously for the same doctor; 2) due to system malfunction or malicious attack, the proxy could send the medical record of another patient or a file of the correct form but carrying wrong information. The consequence of treating the patient based on incorrect information could be very serious or even catastrophic.

The above observation motivates us to study ABE with verifiable outsourced decryption in this paper.

*Our contributions*

In green *et al.* [12] provided the verifiability of the cloud's transformation and provided a method to check the correctness of the transformation. However the authors did not formally define verifiability.

In this paper we first modify the original model of ABE with outsourced decryption in [12] to allow for verifiability of the transformations. After describing the formal definition of verifiability, we propose a new ABE model and based on this new model construct a concrete ABE scheme with verifiable outsourced decryption. Our scheme does not rely on random oracles

The rest of the paper we only focus on CP-ABE with verifiable outsourced decryption. The same approach applies to KP-ABE with verifiable outsourced decryption which will omit here in order to keep the paper compact.

To access the performance of our ABE scheme with verifiability outsourced decryption we implement the CP-ABE scheme with verifiable outsourced decryption and conduct experiments on both an ARM-based mobile device and Intel-core personal computer to model a mobile user and a proxy respectively.

*Literature Survey*

In cloud environments if a data owner wants to share data with users he will encrypt data and then upload to cloud storage service. Through the encryption the cloud cannot know the information of the encrypted data. Besides to avoid the unauthorized user accessing the encrypted data in the cloud, a data owner uses encryption scheme for access control of encrypted data. In existing schemes many encryption schemes can achieve and provide security assure data confidentiality and prevent collusion attack scheme. One of the attribute-based encryption scheme. according to the access policy two types of these schemes can be classified the key-policy and ciphertext-policy attribute-based encryption schemes. The key-policy attribute-based encryption scheme is that the access policy is attached to the user's private key and a set of descriptive attributes is in the encrypted data. If a set of attributes satisfies the access policy the user will recover the message. If not he cannot it

The ciphertext-policy attribute-based scheme is that the access policy is associated to the encrypted data, and a set of descriptive attributes is in the user's private key. If a set attribute satisfies the access policy, the user can decipher the encrypted data.

In the subsection we review some closely related works including non interactive verifiable computation, pairing delegation and proxy reencryption.

*Non-interactive Verifiable Computation:* Non-interactive verifiable computation[19], [20] enables a computationally weak client to outsource the computation of a function to one or more workers. The workers return the result of the function evaluation as well as a non-interactive proof that the computation of the function was carried out correctly. Since these schemes [19], [20] deal with outsourcing of general computation problems and preserve the privacy of input data, they can be used to outsource decryption in ABE systems. However the schemes proposed in [19], [20] use Gentry's fully homomorphic encryption system[21] as a building block, thus the overhead in these schemes is currently too large to be practical [22]. Parno *et al*. [23] establish an important connection between verifiable computation and ABE. They show how to construct a verifiable computation scheme with public delegation and public verifiability from any ABE scheme and how to construct a multifunction verifiable computation scheme form the ABE scheme with outsourced decryption presented in [12]. Goldwasser *et al*. [24] propose a succinct functional encryption scheme, one can obtain a delegation scheme with is both publicly verifiable and secret, in the sense that the prover does not learn anything about input or output of the function being delegated. All these schemes [19], [20], [23], [24] focus on delegating general functions and are not sufficiently efficient for the problem at hand.

*Pairing Delegation:* pairing delegation [25], [26] enables a client to outsource the computation of pairings to another entity. However the schemes proposed in [25], [26] still require the client to compute multiple exponentiations in the target group for every pairing it outsources.

Most importantly when using pairing delegation in the decryption of ABE ciphertexts the amount of computation of the client is still proportional to the size of the access policy. Tsang *et al. [27]* consider batch pairing delegation. However the scheme proposed in [27] can only handle batch delegation for pairings in which one of the points is a constant and it still requires the client to compute a pairing.

*Proxy Reencryption:* in ABE with outsourced decryption a user provides the cloud with a transformation key that allows the cloud to translate an ABE ciphertext on message m into a simple ciphertext on the same m, without learning anything about m. this is reminiscent allows a proxy using a reencryption key to transform an encryption of m. we emphasize that in the model of proxy reencryption, verifiability of the proxy's transformation cannot be achieved.

*Proposed CP-ABE scheme with verifiable outsourced Decryption*

For the national purposes in the below we denote the above CP-ABE scheme as Basic CP-ABE. Based on Basic CP-ABE we present a CP-ABE scheme with verifiable outsourced decryption. The Setup, KeyGen, Encrypt and Decrypt algorithms operate exactly as in Basic CP-ABE. We describe the Reencryption algorithms:

- GenTK$_{out}$(PK, SK$_S$) this algorithm takes as input the public parameters PK and a private key SK$_S$= (S,K, K$_0$, K$_i$) for a set of attributes S. it chooses a random value z$\in$ Z$^*_P$. Then it sets the transformation key as TK$_S$= (S, K$^{'}$=K$^{1/z}$, K$^{'}_0$= K$_0^{1/z}$, K$^{'}_i$=K$_i^{1/z}$) and the retrieving key as RK$_S$=z. note that with overwhelming probability z has multiplicative inverse.

- Tranform$_{out}$(PK, CT, TK$_S$) This algorithm takes as input parameters PK, a ciphertext CT=(A= (A, ρ), C, C$_1$, C$_1^{'}$, C$_{1,I}$ D$_{1,I}$, C$_2$, C$_2^{'}$, C$_{2,I}$, D$_{2,i}$) for an access structure A=(A,ρ), and a transformation key TK$_S$= (S, K$^{'}$, K$^{'}_0$, K$^{'}_i$) for a set of attributes S. it then computes:

$$T_1^{'} = \frac{e(C_1^{'}, K^{'})}{(\Pi_{i\in I}(e(C_{1,I}, K^{'}_0) \cdot E(K^{'}_{\rho(i)}, D_{1,i}))^{w_i}}$$

$$= \frac{e(g,g)^{\alpha s/z} e(g,g)^{ats/z}}{(\Pi_{i\in I} e(g,g)^{atA_i \cdot v \cdot w_i/z})}$$

$$= e(g,g)^{\alpha s/z},$$

$$T_2^{'} = \frac{e(C_2^{'}, K^{'})}{(\Pi_{i\in I}(e(C_{2,I}, K^{'}_0) \cdot e(K^{'}_{\rho(i)}, D_{2,i}))^{w_i}}$$

$$= \frac{e(g,g)^{\alpha s'/z} e(g,g)^{ats'/z}}{(\Pi_{i\in I} e(g,g)^{atA_i \cdot v' \cdot w_i/z})}$$

$$= e(g,g)^{\alpha s'/z},$$

and outputs the transformed ciphertext as CT$^{'}$=(T$^{\wedge}$= C$^{\wedge}$, T$_1$= C$_1$, T$_1^{'}$, T$_2$= C$_2$, T$_2^{'}$).

- Decrypt$_{out}$(PK, CT, CT$^{'}$, RK$_S$) this algorithm takes as input the public parameters PK, a ciphertext CT= (A=(A,ρ), C$^{\wedge}$, C$_1$, C$_1^{'}$, C$_{1,I}$, D$_{1,I}$, C$_2$, C$_2^{'}$, C$_{2,I}$, D$_{2,i}$), a transformed ciphertext CT$^{'}$=(T, T$_1$, T$_1^{'}$, T$_2$, T$_2^{'}$) and a retrieving key RK$_S$=z for a set of attributes S. If T$^{\wedge}$≠C$^{\wedge}$ or T$_1$≠C$_1$ or T2≠C$_2$ it outputs ⊥. Then it computes M= T$_1$/T$_1^{'z}$ and M= T$_2$/T$_2^{'z}$. if T$^{\wedge}$=u$^{H(M)}$v$^{H(M)}$d, it outputs the message M; otherwise it outputs ⊥.

**Security Model for CP-ABE:**

**Setup:** The challenger runs the Setup algorithm and gives the public parameters, PK to the adversary.

**Phase 1:** The adversary makes repeated private keys corresponding to sets of attributes $S_1,\ldots,S_{q1}$.

**Challenge:** The adversary submits two equal length messages $M_0$ and $M_1$. In addition the adversary gives a challenge access structure $A^*$ such that none of the sets $S_1,\ldots,S_{q1}$ from phase 1 satisfy the access structure. The challenger flips a random coin b, and encrypts $M_b$ under $A^*$. the ciphertext $CT^*$ is given to the adversary.

**Phase 2:** phase 1 is repeated with the restriction that none of sets of attributes $S_{q1+1},\ldots,S_q$ satisfy the access structure corresponding to the challenge.

**Guess:** the adversary outputs a guess $b^1$ of b.

The advantage of an adversary A in this game is defined as $Pr[b^1=b]-1/2$. We note that the model can easily be extended to handle chosen-ciphertext attacks by allowing for decryption queries in Phase 1 and Phase 2.

**Security intuition:** As in previous attribute-based encryption schemes the main challenge in designing our scheme was to prevent against attacks from colluding users. Like the scheme of Sahai and Waters [24] our solution randomizes users private keys such that they cannot be combined; however in our solution the secret sharing must be embedded into the ciphertext instead to the private keys. Inorder to decrypt an attacker clearly must recover $e(g,g)^{\alpha s}$. Inorder to do this the attacker must pair C from the ciphertext with the D component from some user's private key. This will reduced in the desired value $e(g,g)^{\alpha s}$, but blinded by some value $e(g,g)^{rs}$.

**Efficiency:** The efficiencies of the key generation and encryption algorithms are both fairly straightforward. The encryption algorithm will require two exponentiations for each leaf in the ciphertext's access tree. The ciphertext size will include two group elements for each leaf. The key generation algorithm requires two exponentiations for every attribute given to the user, and the private key consists of two group elements for every attribute.

**Performance Measurements:**

We now provide some information on the performance achieved by cpabe toolkit. As expected cpabe-keygen runs in time precisely linear in the number of attributes associated with the key it is issuing. The running time of cpabe-enc is also almost perfectly linear with respect to the number of leaf nodes in the access policy. The performance of cpabe-dec is somewhat more interesting. It is slightly more difficult to measure in the absence of a precise application, since the decryption time can depend significantly on the particular access trees and set of attributes involved.

In summary cpabe-keygen and cpabe-enc run in a predictable amount of time based on the number of attributes in a key or leaves in a policy tree.

**Conclusion:**

In this paper we considered a new requirement of ABE with outsourced decryption: efficiency, verifiability. We proposed Concrete ABE scheme with verifiable outsourced decryption and proved it is secure and verifiable. As scheme substantially reduced the computation time required for resource limited devices to recover plaintexts.

**Charan kumar is an P.G scholor in the Department of Computer science & engineering, Sir Vishveshwariah Institute of Science and Technology, Madanapalli.**

**K. Dinesh Kumar is an assistant professor in the Department of Computer science & engineering, Sir Vishveshwariah Institute of Science and Technology, Madanapalli.**

**D. Arun Kumar Reddy is an assistant professor in the Department of Computer science & engineering, Sir Vishveshwariah Institute of Science and Technology, Madanapalli.**